\documentstyle[aps,prl]{revtex}
\begin{document}
\title{Anisotropic Black Holes in Einstein and Brane Gravity}
\author{Sergiu I. Vacaru$^{1}$ and Evghenii Gaburov$^{2}$}
\address{~}
\address{$^1$ Physics Department, CSU Fresno,\ Fresno, CA 93740-8031, USA, \\
and \\
Centro Multidisciplinar de Astrofisica - CENTRA, Departamento de Fisica,\\
Instituto Superior Tecnico, Av. Rovisco Pais 1, Lisboa, 1049-001, Portugal\\
sergiuvacaru@venus.nipne.ro, sergiu$_{-}$vacaru@yahoo.com}
\address{~}
\address{$^2$Department of Physics and Astronomy, University\\
of Leicester, University Road, Leicester, LE1 7RH, UK\\
eg35@leicester.ac.uk}
\maketitle

\begin{abstract}
We consider exact solutions of Einstein equations defining static black
holes parametrized by off--diagonal metrics which by anholonomic mappings
can be equivalently transformed into some diagonal metrics with
coefficients being very similar to those from the Schwarzschild and/or
Reissner-N\"ordstrom solutions with anisotropic renormalizations of
constants. We emphasize that such classes of solutions, for instance,  with
ellipsoidal symmetry of horizons, can be constructed even in  general
relativity theory if off--diagonal metrics and anholonomic frames are
introduced into considerations. Such solutions do not violate the Israel's
uniqueness theorems on static black hole configurations \cite{israel}
because at long radial distances one holds the usual Schwarzschild limit. We
show that anisotropic deformations of  the Reissner-N\"ordstrom metric can
be an exact solution on the brane, re-interpreted as a black hole with an
effective electromagnetic like  charge anisotropically induced and polarized
by higher dimension  gravitational interactions.
\end{abstract}

\pacs{04.50.+h, 04.70.-s }

%%%%%%%%%%%%%%%%%%%%%%%%%%%%%%%%%%%%%%%%%%%%%
%\draft \twocolumn[\hsize\textwidth\columnwidth\hsize\csname
%@twocolumnfalse\endcsname
%\preprint{}

%%%%%%%%%%%%%%%%%%%%%%%%%%%%%%%%%%%%%%%%%%%%%%%%%%%%%%%

%\hfill{hep-th/0108xxx}} %\vskip 2pc
%]

The idea of extra--dimension is gone through a renewal in connection to
string/M--theory \cite{hw} which in low energy limits results in models of
brane gravity and/or high energy physics. It was proven that the matter
fields could be localized on a 3--brane in $1+3+n$ dimensions, while gravity
can propagate in the $n$ extra dimensions which can be large (see, e. g.,
\cite{add}) and even not compact, as in the 5-dimensional (in brief, 5D)
warped space models of Randall and Sundrum \cite{rs} (in brief RS, see also
early versions \cite{a}).

The bulk of solutions of 5D Einstein equations and their reductions to 4D
were constructed by using static diagonal metrics and extensions to
solutions with rotations given with respect to holonomic coordinate frames
of references. On the other hand much attention has been paid to
off--diagonal metrics in higher dimensional gravity beginning the Salam,
Strathee and Petracci work \cite{salam} which proved that including
off--diagonal components in higher dimensional metrics is equivalent to
including of $U(1),SU(2)$ and $SU(3)$ gauge fields. Recently, it was shown
in Ref. \cite{v1} that if we consider off--diagonal metrics which can be
equivalently diagonalized to some corresponding anholonomic frames, the RS
theories become substantially locally anisotropic with variations of
constants on extra dimension coordinate or with anisotropic angular
polarizations of effective 4D constants, induced by higher dimension
gravitational interactions.

If matter on a such anisotropic 3D branes collapses under gravity without
rotating to form a black hole, then the metric on the brane-world should be
close to some anisotropic deformations of the Schwarzschild metric at
astrophysical scales in order to preserve the observationally tested
predictions of general relativity. We emphasize that it is possible to
construct anisotropic deformations of spherical symmetric black hole
solutions to some static configurations with ellipsoidal or toroidal
horizons even in the framework of 4D and in 5D Einstein theory if
off--diagonal metrics and associated anholonomic frames and nonlinear
connections are introduced into consideration \cite{v2}.

Collapse to locally isotropic black holes in the Randall-Sundrum brane-world
scenario was studied by Chamblin et al.~\cite{chr} (see also~\cite{ehm,gkr}
and a review on the subject \cite{maartens}). The item of definition of
black hole solutions have to be reconsidered if we are dealing with
off--diagonal metrics, anholonomic frames both in general relativity and on
anisotropic branes.

In this Letter, we give four classes of exact black hole solutions which
describes ellipsoidal static deformations with anisotropic polarizations and
running of constants of the Schwarzschild and Reissner-N\"ordstrom
solutions. We analyze the conditions when such type anisotropic solutions
defined on 3D branes have their analogous in general relativity.

The 5D pseudo--Riemannian spacetime is provided with local coordinates $%
u^\alpha =(x^i,y^a)=(x^1=f,x^2,x^3,y^4=s,y^5=p),$ where $f$ is the extra
dimension coordinate, $(x^2,x^3)$ are some space coordinates and $(s=\varphi
,p=t)$ (or inversely, $(s=t,p=\varphi )$) are correspondingly some angular
and time like coordinates (or inversely). We suppose that indices run
corresponding values: $i,j,k,...=1,2,3$ and $a,b,c,...=4,5.$ The local
coordinate bases $\partial _\alpha =(\partial _i,\partial _a),$ and their
duals, $d^\alpha =\left( d^i,d^a\right) ,$ are defined respectively as
\begin{equation}
\partial _\alpha \equiv \frac \partial {du^\alpha }=(\partial _i=\frac %
\partial {dx^i},\partial _a=\frac \partial {dy^a}) \mbox{ and } d^\alpha
\equiv du^\alpha =(d^i=dx^i,d^a=dy^a).  \label{pdif}
\end{equation}

For the 5D line element $dl^{2}=G_{\alpha \beta }du^{\alpha }du^{\beta }$ we
choose the metric coefficients $G_{\alpha \beta }$ (with respect to the
coordinate frame (\ref{pdif})) to be parametrized by a off--diagonal matrix
(ansatz) {\small
\begin{equation}
\left[
\begin{array}{ccccc}
1+w_{1}^{\ 2}h_{4}+n_{1}^{\ 2}h_{5} & w_{1}w_{2}h_{4}+n_{1}n_{2}h_{5} &
w_{1}w_{3}h_{4}+n_{1}n_{3}h_{5} & w_{1}h_{4} & n_{1}h_{5} \\
w_{1}w_{2}h_{4}+n_{1}n_{2}h_{5} & g_{2}+w_{2}^{\ 2}h_{4}+n_{2}^{\ 2}h_{5} &
w_{2}w_{3}h_{4}+n_{2}n_{3}h_{5} & w_{2}h_{4} & n_{2}h_{5} \\
w_{1}w_{3}h_{4}+n_{1}n_{3}h_{5} & w_{3}w_{2}h_{4}+n_{2}n_{3}h_{5} &
g_{3}+w_{3}^{\ 2}h_{4}+n_{3}^{\ 2}h_{5} & w_{3}h_{4} & n_{3}h_{5} \\
w_{1}h_{4} & w_{2}h_{4} & w_{3}h_{4} & h_{4} & 0 \\
n_{1}h_{5} & n_{2}h_{5} & n_{3}h_{5} & 0 & h_{5}
\end{array}
\right]   \label{ansatz0}
\end{equation}
} where the coefficients are some necessary smoothly class functions of
type:
\[
g_{2,3}=g_{2,3}(x^{2},x^{3}),h_{4,5}=h_{4,5}(x^{1},x^{2},x^{3},s),w_{i}=w_{i}(x^{1},x^{2},x^{3},s),n_{i}=n_{i}(x^{1},x^{2},x^{3},s).
\]

The line element (\ref{ansatz0}) can be equivalently rewritten in the form
\begin{equation}
\delta l^2=g_{ij}\left( x^2,x^3\right) dx^idx^i+h_{ab}\left(
x^1,x^2,x^3,s\right) \delta y^a\delta y^b,  \label{dmetric}
\end{equation}
with diagonal coefficients $g_{ij}= diag[1,g_2,g_3]$ and $h_{ab}= diag
[h_4,h_5]$ if instead the coordinate bases  (\ref{pdif}) one introduce the
anholonomic frames (anisotropic bases)
\begin{equation}
{\delta }_\alpha \equiv \frac \delta {du^\alpha }=(\delta _i=\partial
_i-N_i^b(u)\ \partial _b,\partial _a=\frac \partial {dy^a}) \mbox{ and }
\delta ^\alpha \equiv \delta u^\alpha = (\delta ^i=dx^i,\delta
^a=dy^a+N_k^a(u)\ dx^k)  \label{ddif}
\end{equation}
where the $N$--coefficients are parametrized $N_i^4=w_i$ and $N_i^5=n_i$
(on anholonomic frame method see details in \cite{v1}).

The nontrivial components of the 5D vacuum Einstein equations, $R_{\alpha
}^{\beta }=0,$ for the ansatz (\ref{dmetric}) given with respect to
anholonomic frames (\ref{ddif}) are
\begin{eqnarray}
R_{2}^{2}=R_{3}^{3}=-\frac{1}{2g_{2}g_{3}}[g_{3}^{\bullet \bullet }-\frac{%
g_{2}^{\bullet }g_{3}^{\bullet }}{2g_{2}}-\frac{(g_{3}^{\bullet })^{2}}{%
2g_{3}}+g_{2}^{^{\prime \prime }}-\frac{g_{2}^{^{\prime }}g_{3}^{^{\prime }}%
}{2g_{3}}-\frac{(g_{2}^{^{\prime }})^{2}}{2g_{2}}] &=&0,  \label{ricci1a} \\
R_{4}^{4}=R_{5}^{5}=-\frac{\beta }{2h_{4}h_{5}} &=&0,  \label{ricci1b} \\
R_{4i}=-w_{i}\frac{\beta }{2h_{5}}-\frac{\alpha _{i}}{2h_{5}} &=&0,
\label{ricci1c} \\
R_{5i}=-\frac{h_{5}}{2h_{4}}\left[ n_{i}^{\ast \ast }+\gamma n_{i}^{\ast }%
\right]  &=&0,  \label{ricci1d}
\end{eqnarray}
where
\[
\alpha _{i}=\partial _{i}{h_{5}^{\ast }}-h_{5}^{\ast }\partial _{i}\ln \sqrt{%
|h_{4}h_{5}|},\beta =h_{5}^{\ast \ast }-h_{5}^{\ast }[\ln \sqrt{|h_{4}h_{5}|}%
]^{\ast },\gamma =(3h_{5}/2h_{4})-h_{4}^{\ast }/h_{4},
\]
the partial derivatives are denoted like $a\symbol{94}=\partial a/\partial
x^{1},h^{\bullet }=\partial h/\partial x^{2},f^{\prime }=\partial f/\partial
x^{2}$ and $f^{\ast }=\partial f/\partial s.$

The system of second order nonlinear partial equations (\ref{ricci1a})--(\ref
{ricci1d}) can be solved in general form:

The equation (\ref{ricci1a}) relates two functions $g_2(x^2,x^3)$ and $%
g_3(x^2,x^3).$ It is solved, for instance, by arbitrary two functions $%
g_2(x^2)$ and $g_3(x^3),$ or by $g_2=g_3=g_{[0]}\exp [a_2x^2+a_3x^3],$ were $%
g_{[0]},a_2$ and $a_3$ are some constants. For a given parametrization of $%
g_2=b_2(x^2)c_2(x^3)$ we can find a decomposition in series for $%
g_3=b_3(x^2)c_3(x^3)$ (in the inverse case a multiple parametrization is
given for $g_3$ and we try to find $g_2);$ for simplicity we omit such
cumbersome formulas. We emphasize that we can always redefine the variables $%
\left( x^2,x^3\right) ,$ or (equivalently) we can perform a 2D conformal
transform to the flat 2D line element
\[
g_2(x^2,x^3)(dx^2)^2+g_3(x^2,x^3)(dx^3)^2\rightarrow (dx^2)^2+(dx^3)^2,
\]
for which the solution of (\ref{ricci1a}) becomes trivial.

The next step is to find solutions of the equation (\ref{ricci1b}) which
relates two functions $h_4\left( x^i,s\right) $ and $h_5\left( x^i,s\right) $%
. This equation is satisfied by arbitrary pairs of coefficients $h_4\left(
x^i,s\right) $ and $h_{5[0]}\left( x^i\right) .$ If dependencies of $h_5$ on
anisotropic variable $s$ are considered, there are two possibilites:

a) to compute
\begin{eqnarray*}
\sqrt{|h_5|} &=&h_{5[1]}\left( x^i\right) +h_{5[2]}\left( x^i\right) \int
\sqrt{|h_4\left( x^i,s\right) |}ds,~h_4^{*}\left( x^i,s\right) \neq 0; \\
&=&h_{5[1]}\left( x^i\right) +h_{5[2]}\left( x^i\right) s,h_4^{*}\left(
x^i,s\right) =0,
\end{eqnarray*}
for some functions $h_{5[1,2]}\left( x^i\right) $ stated by boundary
conditions;

b) or, inversely, to compute $h_4$ for a given $h_5\left( x^i,s\right)
,h_5^{*}\neq 0,$%
\begin{equation}
\sqrt{|h_4|}=h_{[0]}\left( x^i\right) (\sqrt{|h_5\left( x^i,s\right) |})^{*},
\label{p1}
\end{equation}
with $h_{[0]}\left( x^i\right) $ given by boundary conditions.

Having the values of functions $h_4$ and $h_5,$ we can define the
coefficients $w_i\left( x^i,s\right) $ and $n_i\left( x^i,s\right) :$

The exact solutions of (\ref{ricci1c}) is found by solving linear algebraic
equation on $w_k,$
\begin{equation}
w_k=\partial _k\ln [\sqrt{|h_4h_5|}/|h_5^{*}|]/\partial _s\ln [\sqrt{|h_4h_5|%
}/|h_5^{*}|],  \label{w}
\end{equation}
for $\partial _s=\partial /\partial s$ and $h_5^{*}\neq 0.$ If $h_5^{*}=0$
the coefficients $w_k$ could be arbitrary functions on $\left( x^i,s\right)
. $

Integrating two times on variable $s$ we find the exact solution of (\ref
{ricci1d}),
\begin{eqnarray}
n_k &=&n_{k[1]}\left( x^i\right) +n_{k[2]}\left( x^i\right) \int [h_4/(\sqrt{%
|h_5|})^3]ds,~h_5^{*}\neq 0;  \nonumber \\
&=&n_{k[1]}\left( x^i\right) +n_{k[2]}\left( x^i\right) \int
h_4ds,~h_5^{*}=0;  \label{n} \\
&=&n_{k[1]}\left( x^i\right) +n_{k[2]}\left( x^i\right) \int [1/(\sqrt{|h_5|}%
)^3]ds,~h_4^{*}\neq 0,  \nonumber
\end{eqnarray}
for some functions $n_{k[1,2]}\left( x^i\right) $ stated by boundary
conditions.

We shall construct some classes of exact solutions of 5D and 4D vacuum
Einstein equations describing anholonomic deformations of black hole
solutions of the Reissner-N\"{o}rdstrom and Schwarzschild metrics. We
consider two systems of 3D space coorinates:

a) The isotropic spherical coordinates $(\rho ,\theta ,\varphi ),$
\thinspace where the isotropic radial coordinate $\rho $ is related with the
usual radial coordinate $r$ via relation $r=\rho \left( 1+r_g/4\rho \right)
^2$ for $r_g=2G_{[4]}m_0/c^2$ being the 4D gravitational radius of point
particle of mass $m_0,$ $G_{[4]}=1/M_{P[4]}^2$ is the 4D Newton constant
expressed via Plank mass $M_{P[4]}$ which following modern string/brane
theories can considered as a value induced from extra dimensions,  we shall
put the light speed constant $c=1$ (this system of coordinates is
considered, for instance, for the so--called isotropic representation of the
Schwarzschild solution \cite{ll}).

b) The rotation ellipsoid coordinates (in our case isotropic, in brief
re--coordinates) \cite{korn} $(u,v,\varphi )$ with $0\leq u<\infty ,0\leq
v\leq \pi ,0\leq \varphi \leq 2\pi ,$ where $\sigma =\cosh u=4\rho /r_g\geq
1 $ are related with  the isotropic 3D Cartezian coodrinates $(\tilde{x} =
\sinh u\sin v\cos \varphi ,\tilde{y}=\sinh u\sin v\sin \varphi , \tilde{z}
=\cosh u\cos v)$ and define an elongated rotation ellipsoid hypersurface $%
\left( \tilde{x}^2+\tilde{y}^2\right) /(\sigma ^2-1)+\tilde{z}^2/\sigma ^2=1.
$

By straightforward calculations we can verify that we can generate from the
ansatz (\ref{ansatz0}) four classes of exact solutions of the system (\ref
{ricci1a})--(\ref{ricci1d}):

\begin{enumerate}
\item  The anisotropic Reissner-N\"{o}rdstrom black hole solutions with
polarizations on extra dimension and 3D space coordinates  are parametrized
by the data
\begin{eqnarray}
g_2 &=&\left( \frac{1-\frac{r_g}{4\rho }}{1+\frac{r_g}{4\rho }}\right) \frac %
1{\left[ \rho ^2+a\rho /(1+\frac{r_g}{4\rho })^2+b/(1+\frac{r_g}{4\rho })^4%
\right] },g_3=1;  \label{sol1a} \\
h_5 &=&-\frac 1{\rho ^2\left( 1+\frac{r_g}{4\rho }\right) ^4}[1+\frac{%
a\sigma _m\left( f,\rho ,\theta ,\varphi \right) }{\rho \left( 1+\frac{r_g}{%
4\rho }\right) ^2}+\frac{b\sigma _q\left( f,\rho ,\theta ,\varphi \right) }{%
\rho ^2\left( 1+\frac{r_g}{4\rho }\right) ^4}], h_4 =\sin ^2\theta \left[
\left( \sqrt{\left| h_5(f,\rho ,\theta ,\varphi )\right| }\right) \right]
^2\quad \mbox{(see (\ref{p1}))};  \nonumber
\end{eqnarray}
where $a,b$ are constants and $\sigma _m\left( f,\rho ,\theta ,\varphi
\right) $ and $\sigma _q\left( f,\rho ,\theta ,\varphi \right) $ are called
respectively mass and charge polarizations and the coordinates are $\left(
x^i,y^a\right) =\left( f,\rho ,\theta ,t,\varphi \right) .$

\item  The anisotropic Reissner-N\"{o}rdstrom black hole solutions with
extra dimension and time running of constants are parametrized by the data
\begin{eqnarray}
g_2 &=&\left( \frac{1-\frac{r_g}{4\rho }}{1+\frac{r_g}{4\rho }}\right) \frac %
1{\left[ \rho ^2+a\rho /(1+\frac{r_g}{4\rho })^2+b/(1+\frac{r_g}{4\rho })^4%
\right] },g_3=1;  \label{sol1b} \\
h_4 &=&-\frac 1{\rho ^2\left( 1+\frac{r_g}{4\rho }\right) ^4}[1+\frac{%
a\sigma _m\left( f,\rho ,\theta ,t\right) }{\rho \left( 1+\frac{r_g}{4\rho }%
\right) ^2}+\frac{b\sigma _q\left( f,\rho ,\theta ,t\right) }{\rho ^2\left(
1+\frac{r_g}{4\rho }\right) ^4}],h_5=\sin ^2\theta ,  \nonumber
\end{eqnarray}
where $a,b$ are constants and $\sigma _m\left( f,\rho ,\theta ,\varphi
\right) $ and $\sigma _q\left( f,\rho ,\theta ,\varphi \right) $ are called
respectively mass and charge polarizations and the coordinates are $\left(
x^i,y^a\right) =\left( f,\rho ,\theta ,\varphi ,t\right) .$

\item  The ellipsoidal Schwarzschild like black hole solutions with
polarizations on extra dimension and 3D space coordinates  are parametrized
by the data $g_2=g_3 =1$ and
\begin{equation}
h_5 = -\frac{r_g^2}{16}\frac{\cosh ^2u}{\left( 1+\cosh u\right) ^4}\left(
\frac{\cosh u_m(f,u,v,\varphi )-\cosh u}{\cosh u_m(f,u,v,\varphi )+\cosh u}%
\right) ^2,\ h_4 =\frac{\sinh ^2u\sin ^2v}{\sinh ^2u+\sin ^2v}\left[ \left(
\sqrt{\left| h_5(f,u,v,\varphi )\right| }\right) \right] ^2,  \label{sol2a}
\end{equation}
where $\sigma _m=\cosh u_m$ and the coordinates are $\left( x^i,y^a\right)
=\left( f,u,v,\varphi ,t\right) .$

\item  The ellipsoidal Schwarzschild like black hole solutions with extra
dimension and time running of constants are parametrized by the data $%
g_2=g_3=1$ and
\begin{equation}
h_4 = -\frac{r_g^2}{16}\frac{\cosh ^2u}{\left( 1+\cosh u\right) ^4}\left(
\frac{\cosh u_m(f,\rho ,\theta ,t)-\cosh u}{\cosh u_m(f,\rho ,\theta
,t)+\cosh u}\right) ^2, h_5 = \frac{\sinh ^2u\sin ^2v}{\sinh ^2u+\sin ^2v},
\label{sol2b}
\end{equation}
where $\sigma _m=\cosh u_m$ and the coordinates are $\left( x^i,y^a\right)
=\left( f,u,v,t,\varphi \right) .$
\end{enumerate}

The N--coefficients $w_i$ and $n_i$ for the solutions (\ref{sol1a})--(\ref
{sol2b}) are computed respectively following formulas (\ref{w}) and (\ref{n}%
) (we omit the final expressions in this paper).

The mathematical form of the solutions (\ref{sol1a}) and (\ref{sol1b}), with
constants $a=-2m/M_p^2$ and $b=Q,$ is very similar to that of the
Reissner-N\"{o}rdstrom solution from RS gravity \cite{maartens}, but
multiplied on a conformal factor $\left( 1+\frac{r_g}{4\rho }\right)
^{-4}\rho ^{-2},$ with renormalized factors $\sigma _m$ and $\sigma _q$ and
{\em without electric charge} being present. The induced 4D gravitational
''receptivities'' $\sigma _m$ and $\sigma _q$ in (\ref{sol1a}) emphasize
dependencies on coordinates $\left( f,\rho ,\theta ,\varphi \right) ,$ where
$s=\varphi $ is the anisotropic coordinate. In a similar fashion one induces
running on time and the 5th coordinate, and anisotropic polarizations on $%
\rho $ and $\theta ,$ of constants for the solution (\ref{sol1b}).

Instead the Reissner-N\"{o}rdstrom-type correction to the Schwarzschild
potential the mentioned polarizations can be thought as defined by some
nonlinear higher dimension gravitational interactions and anholonomic frame
constraints for anisotropic Reissner-N\"{o}rdstrom black hole configurations
with a {\em `tidal charge'} $Q$ arising from the projection onto the brane
of free gravitational field effects in the bulk. These effects are
transmitted via the bulk Weyl tensor, off--diagonal components of the metric
and by anholonomic frames. The Schwarzschild potential $\Phi =-M/(M_{{\rm p}%
}^2r)$, where $M_{{\rm p}}$ is the effective Planck mass on the brane, is
modified to
\begin{equation}
\Phi =-{\frac{M\sigma _m}{M_{{\rm p}}^2r}}+{\frac{Q\sigma _q}{2r^2}}\,,
\label{npot}
\end{equation}
where the `tidal charge' parameter $Q$ may be positive or negative. The
possibility to modify anisotropically the Newton law via effective
anisotropic masses $M\sigma _m,$ or by anisotropic effective 4D Plank
constants, renormalized like $\sigma _m/M_{{\rm p}}^2,$ was recently
emphasized in Ref. \cite{v1}. In this paper we state that there are possible
additional renormalizations of the ''effective'' electric charge, $Q\sigma
_q.$ For diagonal metrics we put $\sigma _m=\sigma _q=1$  and by
multiplication on corresponding conformal factors and  with respect to
holonomic frames we recover the locally isotropic results from Refs. \cite
{maartens}. We must also impose the condition that the 5D spacetime is
modeled as a $AdS_5$ slice provided with an anholonomic frame structure.

The renormalized tidal charge $Q\sigma _q$ affects the geodesics and the
gravitational potential, so that indirect limits may be placed on it by
observations. Nevertheless, current observational limits on $|Q\sigma _q|$
are rather weak, since the correction term in Eq.~(\ref{npot}) decreases off
rapidly with increasing $r$, and astrophysical measurements (lensing and
perihelion precession) probe mostly (weak-field) solar scales.

Now we analyze the properties of solutions (\ref{sol2a}) and (\ref{sol2b}).
They describe Schwarzschild like solutions with the horizon forming a
rotation ellipsoid horizon. For the general relativity such solutions were
constructed in Refs. \cite{v2}. Here, it should be emphasized that static
anisotropic deformations of the Schwarzschild metric are described by
off--diagonal metrics and corresponding conformal transforms. At large
radial distances from the horizon the anisotropic configurations transform
into the usual one with spherical symmetry. That why the solutions with
anisotropic rotation ellipsoidal horizons do not contradict the well known
Israel and Carter theorems \cite{israel} which were proved in the assumption
of spherical symmetry at asymptotics. Anisotropic 4D black hole solutions
follow from the data (\ref{sol2a}) and (\ref{sol2b}) if you state some
polarizations depending only on 3D space coordinates ($u,v,\varphi ),$ or on
some of them. In this paper we show that in 5D there are warped to 4D static
ellipsoidal like solutions with constants renormalized anisotropically on
some 3D space coordinates and on extra dimension coordinate (in the class of
solutions (\ref{sol2a})) and running of constants on time and the 5th
coordinate, with possible additional polarizations on some 3D coordinates
(in the class of solutions (\ref{sol2b})).

A geometric approach to the Randall-Sundrum scenario has been developed by
Shiromizu et al.~\cite{sms} (see also~\cite{bdul}), and proves to be a
useful starting point for formulating the problem and seeing clear lines of
approach. In this work we considered a variant of anholonomic RS geometry.
The vacuum solutions (\ref{sol1a})--(\ref{sol2b}) localized on the brane
must satisfy the 5D equation in the Shiromizu et al. representation if in 4D
some sources are considered as to be induced from extra dimension gravity.

The method of anholonomic frames covers the results on linear
extensionss of the Schwarzschild horizon into the bulk
\cite{gian}. The solutions presented in this paper are
nonlinearly induced, are based on very general method of
construction exact solutions in extra dimension gravity and
generalize also the Reissner-N\"{o}rdstrom solution from RS
gravity. The obtained solutions are locally anisotopic but,
nevertheless, they posses local 4D Lorentz symmetry, which is
explicitly emphasized with respect to anholonomic frames. There
are possible constructions with broken Lorentz symmetry as in
\cite {csaki} (if we impose not a locally isotropic limit of our
solutions, but an anisotropic static one). We omit such
considerations here.

In conclusion we formulate a prescription for mapping 4D general relativity
solutions with diagonal metrics to 4D and 5D solutions of brane world: {\em %
a general relativity vacuum solution gives rise to a vacuum brane-world
solution in 5D gravity given with similar coefficients of metrics  but
defined with respect to some anholonomic frames and with  anisotropic
renormalization of fundamental constants; such type of  solutions are
parametrized by off--diagonal metrics if of type  (\ref{ansatz0}) if they
are re--defined with respect to  coordinate frames }.

\vskip3pt {\bf Acknowledgements:} The authors thank D.\ Gontsa for
discussion and collaboration. S. V. is grateful to P. Stavrinos
and D. Singleton for support and hospitality. The work is
supported both by ''The 2000--2001 California State University
Legislative Award'' and a NATO/Portugal fellowship grant at the
Instituto Superior Tecnico,  Lisbon.

%\end{thebibliography}

\end{document}